\documentclass[twocolumn]{jpsj3}
\usepackage{txfonts}
\usepackage{bm}
\usepackage{color}

\title{Multi-Orbital Superconductivity in SrTiO$_{3}$/LaAlO$_{3}$ Interface and SrTiO$_3$ Surface} 

\author{ Yasuharu NAKAMURA$^1$ and Youichi YANASE$^{1,2}$\thanks{E-mail address: yanase@phys.sc.niigata-u.ac.jp}}
\inst{$^1$Graduate School of Science and Technology, Niigata University, Niigata 950-2181, Japan \\
$^2$Department of Physics, Niigata University, Niigata 950-2181, Japan} 

\recdate{April 23, 2013; accepted May 28, 2013}

\abst{
We investigate the superconductivity in two-dimensional electron systems formed in SrTiO$_3$ 
nanostructures. Our theoretical analysis is based on the three-orbital model, which takes into 
account $t_{\rm 2g}$ orbitals of Ti ions. 
Because of the interfacial breaking of mirror symmetry, 
a Rashba-type antisymmetric spin-orbit coupling arises from the cooperation of 
intersite and interorbital hybridyzation and atomic LS coupling. 
This model shows a characteristic spin texture and carrier density dependence of 
Rashba spin-orbit coupling through the orbital degree of freedom. 
Superconductivity is mainly caused by heavy quasiparticles consisting 
of d$_{\rm yz}$ and d$_{\rm zx}$ orbitals at high carrier densities.
We find that the Rashba spin-orbit coupling stabilizes a quasi-one-dimensional 
superconducting phase caused by one of the d$_{\rm yz}$ or d$_{\rm zx}$ orbitals at high magnetic fields 
along interfaces. 
This quasi-one-dimensional superconducting phase is protected against paramagnetic 
depairing effects by the Rashba spin-orbit coupling 
and realizes a large upper critical field $H_{\rm c2}$ beyond  the Pauli-Clogston-Chandrasekhar limit. 
This finding is consistent with an extraordinarily large upper critical field 
observed in SrTiO$_3$/LaAlO$_{3}$ interfaces and its carrier density dependence. 
The possible coexistence of superconductivity and ferromagnetism in SrTiO$_{3}$/LaAlO$_{3}$ 
interfaces may also be attributed to this quasi-one-dimensional superconducting phase. 
}

\kword{Non-centrosymmetric superconductivity, two-dimensional electron gas, multi-orbital model}

\begin{document}
\maketitle

\newcommand{\g}{\mbox{\boldmath$g$}}
\renewcommand{\k}{\mbox{\boldmath$k$}}
\newcommand{\LL}{\mbox{\boldmath$L$}}
\renewcommand{\SS}{\mbox{\boldmath$S$}}
\newcommand{\etal}{{\it et al.}} 


Two-dimensional conducting electron systems formed on SrTiO$_3$ heterostructures 
are attracting much attention. For instance, electron gases with 
a high carrier density on the order of $10^{13}$ ${\rm cm}^{-2}$
have been realized in SrTiO$_3$/LaAlO$_3$ (STO/LAO) interfaces~\cite{Ohtomo}, 
SrTiO$_3$/LaTiO$_3$ interfaces~\cite{Biscaras},
SrTiO$_3$ (STO) surfaces~\cite{Ueno}, and $\delta$-doped STO.~\cite{Kozuka} 
The discovery of superconductivity~\cite{Reyen}, 
ferromagnetism~\cite{Brinkman,Dikin,Li,Bert,Ariando},  
and their coexistence~\cite{Dikin,Li,Bert,Ariando} 
shed light on innovating phenomena in these systems. 
These quantum condensed phases are controlled by a gate voltage through the change of 
carrier density~\cite{Ueno,Caviglia,Bell,Shalom,Caviglia-2}. 
One of the key issues is the role of Rashba-type antisymmetric spin-orbit coupling~\cite{Michaeli} 
arising from the interfacial breaking of mirror symmetry, which may realize an exotic quantum 
condensed phase, such as non-centrosymmetric superconductivity~\cite{Springer}, 
chiral magnetism~\cite{Banerjee}, and their coexistent phase. 
In this research, we theoretically study the non-centrosymmetric superconductivity realized 
in STO nanostructures from the microscopic point of view.


It has been shown that a two-dimensional electron gas is confined in a few TiO$_2$ layers of 
the STO/LAO interface and STO surface in the high-carrier-density 
region~\cite{Ueno,Santander-Sryo,Joshua,Popovic,Pentcheva,Khalsa,Delugas}. 
The conduction bands mainly consist of three t$_{\rm 2g}$ orbitals of Ti 
ions~\cite{Santander-Sryo,Joshua,Popovic,Pentcheva,Khalsa,Delugas}. 
Although the degeneracy of t$_{\rm 2g}$ orbitals significantly affects the band structure 
of two-dimensional electron gases,  a theory of superconductivity based on the 
multi-orbital model has not been conducted. 
Multiband models have been studied~\cite{Mizohata,Fernandes}, 
but the symmetry of t$_{\rm 2g}$ orbitals is taken into 
account in this study for the first time. 
We show that the synergy of broken inversion symmetry and orbital degeneracy 
stabilizes an intriguing superconducting phase in the two-dimensional electron gases.

Our study is based on a two-dimensional tight-binding model that reproduces the 
electronic structure of the STO/LAO interface indicated by first principles band structure 
calculations~\cite{Popovic,Pentcheva,Delugas,Khalsa,Hirayama,Zhong,Khalsa2} 
and experiments~\cite{Santander-Sryo,Joshua}. 
We here focus on the STO/LAO interface, which has been intensively investigated, but our main  
results are also valid for other STO heterostructures.  
The model is described as 
\begin{align}
\hspace*{-20mm}
H=H_{\rm 0} + H_{\rm I} + H_{\rm Z}, 
\end{align}
where the single-particle Hamiltonian $H_{\rm 0}$ is 
\begin{eqnarray}
\hspace*{-11mm} &&
H_{\rm 0} = H_{\rm kin}+H_{\rm hyb}+H_{\rm CEF}+H_{\rm odd}+H_{\rm LS}, \\
\hspace*{-11mm} &&
H_{\rm kin} = \sum_{{\bm k}}\sum_{m=1,2,3}\sum_{s=\uparrow,\downarrow}(\varepsilon_{m}({\bm k})-\mu)
c^{\dagger}_{{\bm k}, \, ms}c_{{\bm k}, \, ms},\\
\hspace*{-11mm} &&
H_{\rm hyb} = \sum_{{\bm k}}\sum_{s=\uparrow,\downarrow}[V({\bm k})c^{\dagger}_{{\bm k}, \, 1s}c_{{\bm k}, \, 2s} 
+ {\rm h.c.}],\\
\hspace*{-11mm} &&
H_{\rm CEF} = \Delta \sum_{i}n_{3i},\\
\hspace*{-11mm} &&
H_{\rm odd} = \sum_{{\bm k}}\sum_{s=\uparrow,\downarrow}
[V_{\rm x}({\bm k})c^{\dagger}_{{\bm k},\, 1s}c_{{\bm k},\, 3s} 
+ V_{\rm y}({\bm k})c^{\dagger}_{{\bm k}, \, 2s}c_{{\bm k}, \, 3s} + {\rm h.c.}],\\
\hspace*{-11mm} &&
H_{\rm LS} = \lambda\sum_{i}{\bm L}_{i} \cdot {\bm S}_{i}. 
\end{eqnarray} 
We denote (d$_{yz}$, d$_{zx}$, d$_{xy}$) orbitals using the index $m=(1,2,3)$, respectively.
The first term $H_{\rm kin}$ describes the kinetic energy of each orbital and includes the chemical potential 
$\mu$. $H_{\rm hyb}$ is the intersite hybridization term of d$_{\rm yz}$ and d$_{\rm zx}$ orbitals.  
$H_{\rm CEF}$ represents the crystal electric field of tetragonal systems. 
Because the mirror symmetry is broken near the interface/surface, hybridization is allowed between 
d$_{\rm xy}$ and d$_{\rm yz}$/d$_{\rm zx}$ orbitals, and is represented by the ``odd parity 
hybridization term'' $H_{\rm odd}$. 
The atomic spin-orbit coupling term (LS coupling term) of Ti ions is taken into 
account in $H_{\rm LS}$.
We here adopt the tight-binding model reproducing first principles band structure 
calculations for STO heterostructures~\cite{Hirayama,Zhong,Khalsa2},  
$\varepsilon_{\rm 1}({\bm k})=-2t_{\rm 3}\cos k_{\rm x}-2t_{\rm 2}\cos k_{\rm y}$, 
$\varepsilon_{\rm 2}({\bm k})=-2t_{\rm 2}\cos k_{\rm x}-2t_{\rm 3}\cos k_{\rm y}$,
$\varepsilon_{\rm 3}({\bm k})=-2t_{\rm 1}(\cos k_{\rm x}+\cos k_{\rm y})-4t_{\rm 4}\cos k_{\rm x}\cos k_{\rm y}$, 
$V({\bm k})=4t_{\rm 5}\sin k_{\rm x}\sin k_{\rm y}$, 
$V_{\rm x}({\bm k})=2{\rm i}t_{\rm odd}\sin k_{\rm x}$, and 
$V_{\rm y}({\bm k})=2{\rm i}t_{\rm odd}\sin k_{\rm y}$. 
The same tight-binding model has been adopted for the study of surface spin-triplet superconductivity in 
Sr$_2$RuO$_4$~\cite{YanaseSr2RuO4_001}. 
Recent studies have examined the Rashba-type antisymmetric spin-orbit coupling~\cite{Zhong,Khalsa2,Nayak} 
and magnetotransport~\cite{Nayak} in STO/LAO interfaces on the basis of this model.

In this paper, we focus on the role of Rashba-type antisymmetric spin-orbit coupling 
in the interface superconductivity. 
In the above model, the Rashba spin-orbit coupling is induced by the combination of 
the odd parity hybridization term, $H_{\rm odd}$, and the LS coupling term, $H_{\rm LS}$. 
The former arises from the parity mixing of local orbitals, which is a general source of 
antisymmetric spin-orbit coupling~\cite{YanaseCePt3Si,Nagano-Shishidou-Oguchi}. 
For instance, the $V_{\rm x}({\bm k})$ ($V_{\rm y}({\bm k})$) term describes 
the mixing of d$_{\rm yz}$ (d$_{\rm zx}$) and d$_{\rm xy}$ orbitals of Ti ions, 
which mainly occurs through the parity mixing with the p$_{\rm y}$ orbital 
(p$_{\rm x}$ orbital) on oxygen ions.

We consider the $s$-wave superconductivity as expected in the bulk STO~\cite{Binnig}. 
Unconventional pairing due to the electron correlation has been studied,~\cite{Yada} 
however, we do not touch this possibility.  
Our reasonable assumption has been justified by the recent experiment on superfluid density~\cite{Bert2}. 
For simplicity, we take into account the intraorbital attractive interaction $U <0$ 
and the interorbital attractive interaction $U' < 0$ in the spin-singlet channel;   
\begin{align}
H_{\rm I}=&U\sum_{i}\sum_{m} n_{i,m\uparrow}n_{i,m\downarrow}+U'\sum_{i}\sum_{m\neq m'}n_{i,m\uparrow}n_{i,m'\downarrow}. 
\end{align} 
For the discussion of the superconducting state in the magnetic field, we consider the Zeeman coupling term
\begin{align}
H_{\rm Z}=&-\sum_{{\bm k}}\sum_{m}\sum_{s,s'}\mu_{\rm B}{\bm H}\cdot{\bm \sigma}_{ss'}
c^{\dagger}_{{\bm k}, \, ms} c_{{\bm k}, \, ms'}, 
\end{align} 
in which ${\bm \sigma}$ is the Pauli matrix and $\mu_{\rm B}$ is the Bohr magneton. 
The orbital depairing effect arising from the coupling of electron motion and vector potential 
is suppressed by the geometry when we consider the magnetic field 
parallel to the two-dimensional conducting plane, ${\bm H} \parallel \hat{x}$. 
The orbital polarization due to the magnetic field is also ignored since the orbital moment 
along the plane vanishes for the degenerate d$_{\rm yz}$/d$_{\rm zx}$ orbitals.


Now, we formulate the linearized gap equation, by which we determine the instability to the 
superconducting phase. 
First, we diagonalize the noninteracting Hamiltonian ($H_{\rm 0} + H_{\rm Z}$) 
using the unitary matrix    
$\hat{U}({\bm k}) = \left(u_{ms,j}({\bm k})\right)$. 
Thereby, the basis changes as $C^{\dagger}_{{\bm k}}=\Gamma^{\dagger}_{{\bm k}}U^{\dagger}({\bm k})$, 
where $C^{\dagger}_{{\bm k}}=(c^{\dagger}_{{\bm k}, \, 1\uparrow},c^{\dagger}_{{\bm k}, \, 2\uparrow},\cdots ,
c^{\dagger}_{{\bm k}, \, 3\downarrow})$ and $\Gamma^{\dagger}_{{\bm k}}=(\gamma^{\dagger}_{{\bm k}, \, 1},
\gamma^{\dagger}_{{\bm k}, \, 2},\cdots ,\gamma^{\dagger}_{{\bm k}, \, 6})$. 
With the use of the operators of quasiparticles, 
$\gamma^{\dagger}_{{\bm k}, j}$ and $\gamma_{{\bm k}, j}$, the noninteracting Hamiltonian is described as,
\begin{eqnarray}
H_{\rm 0}+H_{\rm z}=\sum_{{\bm k}} \sum_{j=1}^{6} E_{j}({\bm k}) \ \gamma^{\dagger}_{{\bm k}, \, j} \ \gamma_{{\bm k}, \, j},
\end{eqnarray}
where $E_{j}({\bm k})$ is a quasiparticle's energy and $E_{i}({\bm k}) \geq E_{j}({\bm k})$ for $i > j$.

Next, we introduce Matsubara Green functions in the orbital basis, 
\begin{align}
G_{m's', \ ms}({\bm k},{\rm i}\omega_{l})&=\int_{0}^{\beta}d\tau e^{{\rm i}\omega_{l}\tau}\langle 
c_{{\bm k}, \, m's'}(\tau)c^{\dagger}_{{\bm k}, \, ms}(0) \rangle, \\
=&\sum^{6}_{j=1}\frac{1}{{\rm i}\omega_{l}-E_{j}({\bm k})}u_{m's', j}({\bm k})u^{*}_{ms, j}({\bm k}), 
\end{align}
where $\omega_{l}$ is the Matsubara frequency. 
The linearized gap equation is obtained by looking at the divergence of 
the T-matrix, $\hat{T}({\bm q})$, which is given by 
\begin{eqnarray}
&& \hspace*{-15mm}
\hat{T}({\bm q}) = \hat{T}_{\rm 0}({\bm q}) - \hat{T}({\bm q}) \hat{H}_{\rm I} \hat{T}_{\rm 0}({\bm q}).   
\end{eqnarray}
The wave vector ${\bm q}$ represents the total momentum of Cooper pairs. 
In our model, the matrix element of the irreducible T-matrix $\hat{T}_{0}({\bm q})$ 
is obtained as 
\begin{eqnarray}
&& \hspace{-14mm}
T_{\rm 0}^{(m n, \, m' n')}({\bm q}) 
\nonumber \\ &&  \hspace{-14mm}
=T \sum_{\omega_{l}}\sum_{{\bm k}}
[ G_{m\uparrow, \, m'\uparrow}({\bm q}/2+{\bm k},{\rm i}\omega_{l}) 
G_{n\downarrow, \, n'\downarrow}({\bm q}/2-{\bm k},-{\rm i}\omega_{l})    
\nonumber \\
&& 
-G_{m\uparrow, \, n'\downarrow}({\bm q}/2+{\bm k},{\rm i}\omega_{l}) 
G_{n\downarrow, \, m'\uparrow}({\bm q}/2-{\bm k},-{\rm i}\omega_{l}) ], 
\end{eqnarray}
where $T$ is the temperature. 
When we represent the T-matrix using the basis $(mn)=(11,12,13,21,22,23,31,32,33)$, 
the interaction term is represented by the $9 \times 9$ diagonal matrix,  
$\hat{H}_{\rm I} = \left( U_{m} \delta_{mn} \right)$ with $U_{m} = U$ for $m=1, 5, 9$ and 
$U_{m}=U'$ for others. 
The superconducting transition occurs when the maximum eigenvalue of the matrix, 
$ - \hat{H}_{\rm I} \hat{T}_{\rm 0}$, is unity. 
Then, an element of the eigenvector 
$(\psi_{mn})$ is proportional to the order parameter  
$\Delta_{mn} = - g \sum_{k} \langle c_{{\bm k}, \, m \uparrow} c_{-{\bm k}, \, n \downarrow} \rangle$, where 
$g=U$ for $m=n$ and $g=U'$ for $m \ne n$. 
In what follows, we assume a zero total momentum of Cooper pairs, namely, ${\bm q}=0$. 
Although a helical superconducting state with ${\bm q} \ne 0$ is stabilized in 
non-centrosymmetric superconductors under the magnetic field,~\cite{Springer,Aoyama} 
a finite momentum ${\bm q}$ does not play any important role in the following results. 
This is because the paramagnetic depairing effect is suppressed by the orbital degree of freedom, 
as we show below.


%
\begin{figure}[ht]
\centering
\includegraphics[width=8.5cm,origin=c,keepaspectratio,clip]{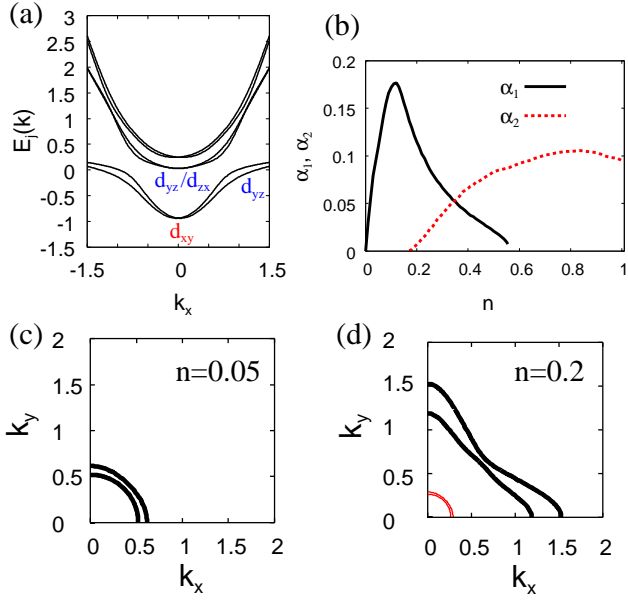}
\caption{(Color online) 
(a) Band structure of our model. We show the dispersion relation 
$E_{j}({{\bm k}})$ for ${\bm k} = (k_{\rm x},0)$. 
The origin is the chemical potential $\mu$ for a carrier density $n=0.15$. 
(b) Carrier density dependence of spin-orbit coupling on the Fermi surface. We show 
$\alpha_1 = E_2({\bm k}_{{\rm F}2}) - E_1({\bm k}_{{\rm F}2})$ (solid line) and 
$\alpha_2 = E_4({\bm k}_{{\rm F}4}) - E_3({\bm k}_{{\rm F}4})$ (dashed line) with ${\bm k}_{{\rm F}j}$ 
being the Fermi wave number of the $j$-th band along the [100] axis. 
(c) and (d) show Fermi surfaces for $n=0.05$ and for $n=0.2$, respectively. 
Other parameters are assumed as 
$(t_{\rm 1}, t_{\rm 2}, t_{\rm 3}, t_{\rm 4}, t_{\rm 5}, \Delta, t_{\rm odd}, \lambda)
=(1.0, 1.0, 0.05, 0.4, 0.1, 2.45, 0.25,0.2)$. 
}
\label{Fig.2}
\end{figure}

We choose the parameters 
\begin{eqnarray}
(t_{\rm 1}, t_{\rm 2}, t_{\rm 3}, t_{\rm 4}, t_{\rm 5}, \Delta)=(1.0, 1.0, 0.05, 0.4, 0.1, 2.45), 
\end{eqnarray} 
so as to reproduce the electronic structure of two-dimensional electron 
gases.~\cite{Santander-Sryo,Joshua,Popovic,Pentcheva,Khalsa,Khalsa2,Delugas,Hirayama,Zhong} 
We choose the unit of energy as $t_1 =1$. Band structure calculations resulted in 
$t_1 = 300$ meV~\cite{Hirayama}, giving rise to an anisotropic Fermi velocity,  
$v_{\rm F} = 7 \times 10^{4} -  4 \times 10^{5}$ ${\rm m/s}$, for $n=0.15$.
For the parameters in eq.~(15), 
the d$_{\rm xy}$ orbital has a lower energy than the d$_{\rm yz}$/d$_{\rm zx}$ orbitals, as expected in STO 
heterostructures~\cite{Santander-Sryo,Joshua,Popovic,Pentcheva,Khalsa,Khalsa2,Delugas,Hirayama,Zhong}; 
the level splitting at the $\Gamma$ point is $-2t_2 -2t_3 + 4t_1 +4 t_4 - \Delta =1.05 \sim 300$meV. 
The chemical potential $\mu$ is determined so that the mobile carrier density per Ti ion is $n$. 
Although an enormous carrier density of $3.5 \times 10^{14}$ ${\rm cm}^{-2}$ corresponding to 
$n=0.5$ at the STO/LAO interface was predicted by the 
``polar catastrophe" mechanism,~\cite{Ohtomo} recent experiments have shown a rather low density of 
mobile carriers~\cite{Caviglia,Bell,Shalom,Caviglia-2,Santander-Sryo,Joshua}. 
One of our purposes is to clarify the carrier density dependence of the superconducting state. 
The sources of Rashba spin-orbit coupling are assumed to be $t_{\rm odd}=0.25$ and $\lambda=0.2$ 
unless mentioned otherwise explicitly. 
We here assume rather large values of $t_{\rm odd}$ and $\lambda$ so that the amplitude of 
Rashba spin-orbit coupling $\alpha \sim t_{\rm odd} \lambda /\Delta$ is larger than 
the transition temperature of superconductivity. 
We assume attractive interactions $U = U'$ so that the transition temperature 
at zero magnetic field is 
$T_{\rm c} = 0.005 = 17$ K. 
A large transition temperature compared with the experimental $T_{\rm c} = 0.3$ K 
is assumed for the accuracy of numerical calculation. Since we discuss the normalized  
$\mu_{\rm B}H_{\rm c2}/T_{\rm c}$, the following results are hardly altered by the magnitude of $T_{\rm c}$.
As we show elsewhere, the superconducting phase is almost independent of the ratio $U'/U$.

Figure~1(a) shows the band structure of our model. 
We see the spin splitting caused by the Rashba spin-orbit coupling. 
Because the Rashba spin-orbit coupling is enhanced around the band 
crossing points~\cite{YanaseCePt3Si}, the magnitude of spin splitting shows a 
nonmonotonic carrier density dependence. 
Figure~1(b) shows the spin splitting in the lowest pair of bands 
[$\alpha_1 = E_2({\bm k}_{{\rm F}2}) - E_1({\bm k}_{{\rm F}2})$] 
and that in the second lowest pair of bands 
[$\alpha_2 = E_4({\bm k}_{{\rm F}4}) - E_3({\bm k}_{{\rm F}4})$] as a function of carrier density, 
where ${\bm k}_{{\rm F}j}$ is the Fermi wave number of the $j$-th band along the [100] axis. 
The nonmonotonic behavior of a spin splitting, $\alpha_1$, is consistent with experimental observations 
for STO/LAO interfaces. The seemingly contradictory carrier density 
dependence~\cite{Shalom,Caviglia-2} of Rashba spin-orbit coupling is probably caused by the peak of 
$\alpha_1$, as pointed out by Zhong \etal~\cite{Zhong}.  
In our model, the Fermi level crosses the bottom of the second lowest pair of bands 
[$E_4(0) = E_3(0) = 0$] at approximately $n = 0.16$. 
The Fermi surfaces for $n = 0.05$ and $n = 0.2$ are shown in Figs.~1(c) and 1(d), 
respectively. The isotropic Fermi surfaces mainly consist of the d$_{\rm xy}$ orbital for 
a low carrier density, $n = 0.05$, while large anisotropic Fermi surfaces mainly consist of 
the d$_{\rm yz}$/d$_{\rm zx}$ orbitals for a large carrier density, $n = 0.2$.

First, we discuss the superconducting state at zero magnetic field. While the superconductivity is 
mainly caused by the d$_{\rm xy}$ orbital at low carrier densities, $n < 0.078$, the intraorbital 
Cooper pairing of d$_{\rm yz}$ and d$_{\rm zx}$ orbitals is the main source of superconductivity 
at high carrier densities, $n > 0.078$. This crossover of the superconducting state coincides with the 
change of quasiparticles on the Fermi surfaces discussed for Figs.~1(c) and 1(d). 
When we assume the attractive interactions $U=U'$ independent of carrier density, 
the transition temperature monotonically increases with increasing carrier density. 
The nonmonotonic carrier density dependence observed in experiments~\cite{Caviglia} is reproduced by assuming 
a decreasing function of $U=U'$ against carrier density. In this study, 
We avoid such a phenomenological assumption and discuss 
the normalized values such as $\mu_{\rm B} H_{\rm c2}/T_{\rm c}$. 
Note that the odd-parity hybridization $t_{\rm odd}$ and LS coupling $\lambda$ hardly affect 
the superconducting state at zero magnetic field.

\begin{figure}[ht]
\centering
\includegraphics[width=7.5cm,origin=c,keepaspectratio,clip]{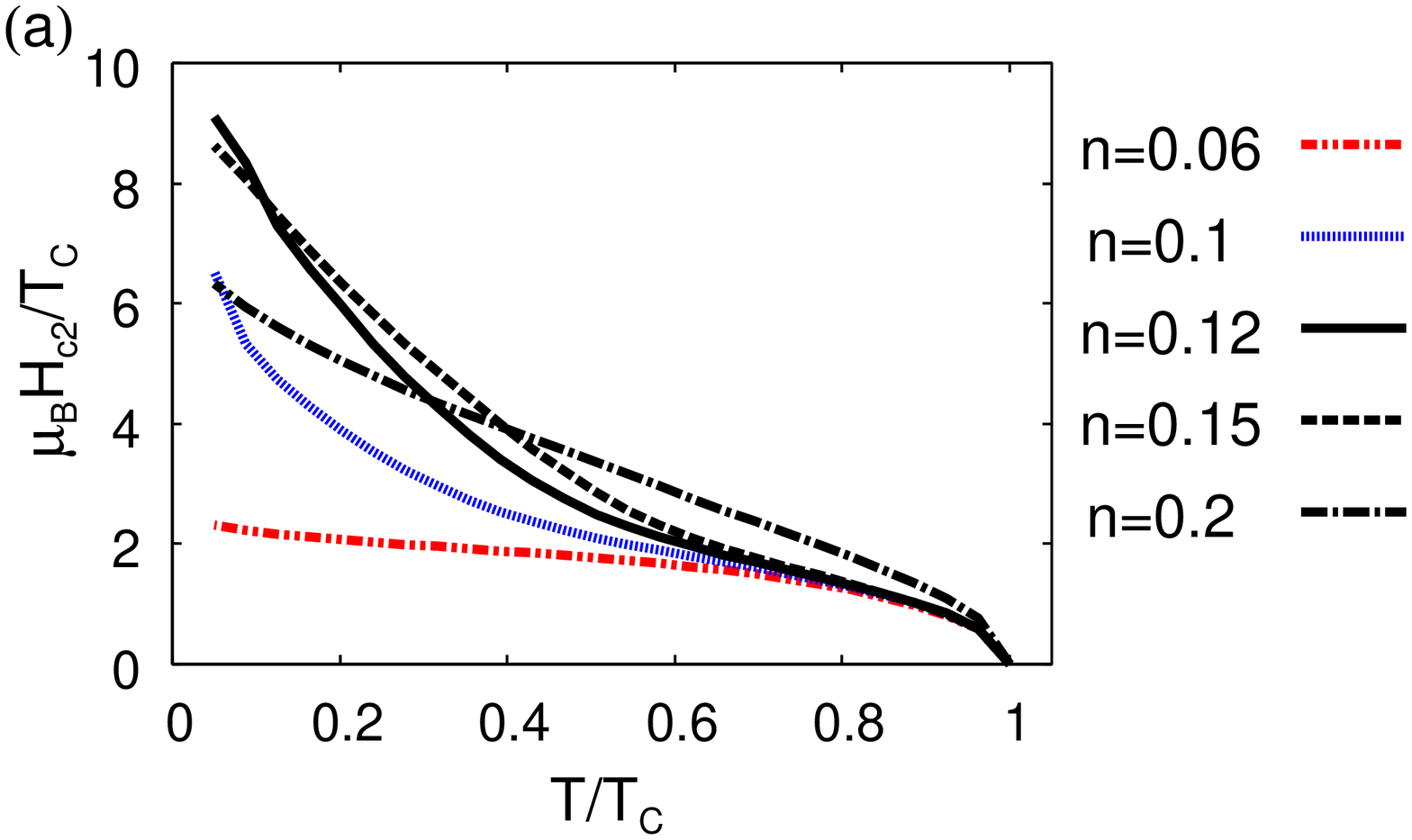}
\hspace*{5mm}
\includegraphics[width=8.3cm,origin=c,keepaspectratio,clip]{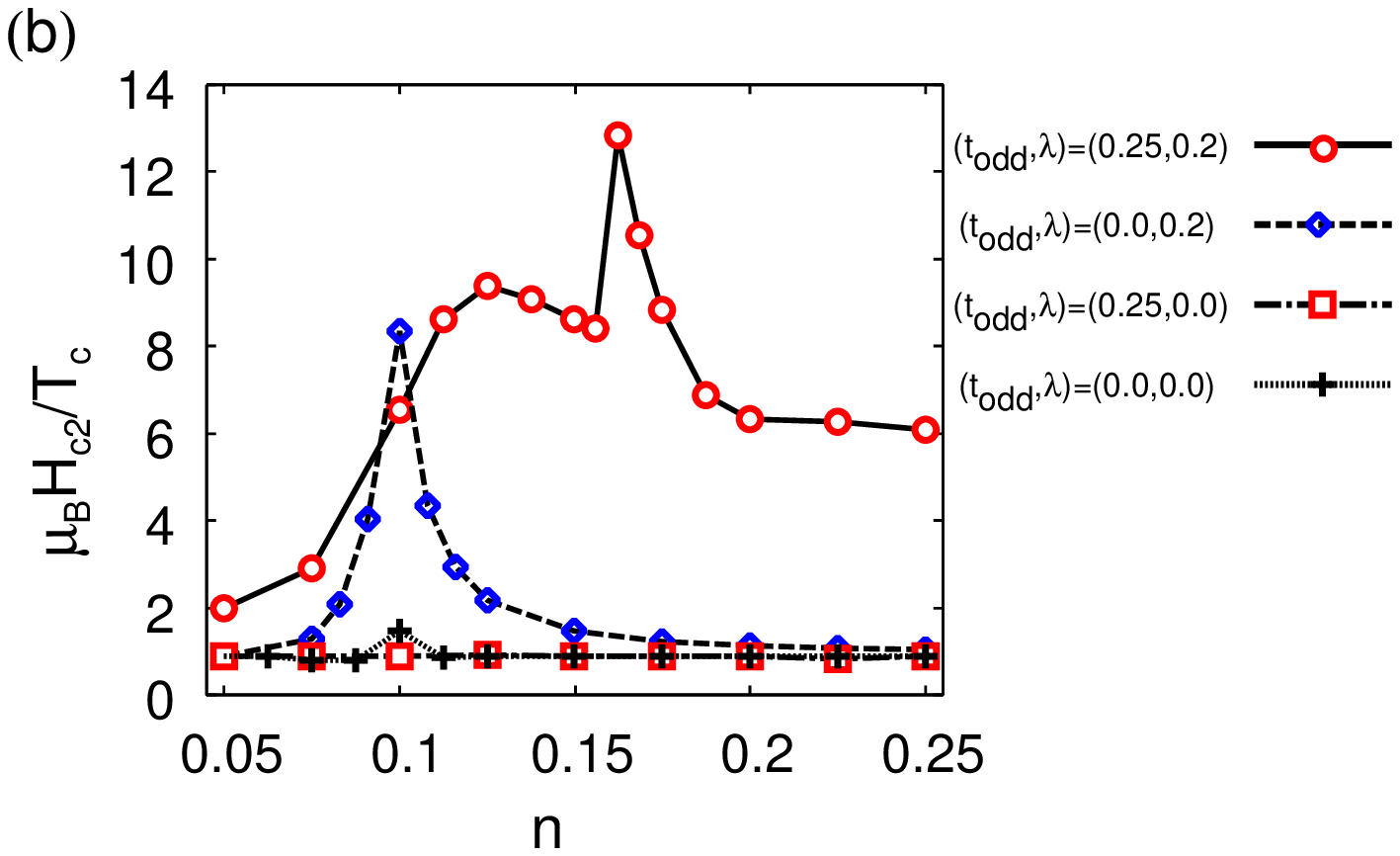}
\hspace{0.0cm}
\caption{(Color online) 
(a) Normalized upper critical field, $\mu_{\rm B} H_{\rm c2}/T_{\rm c}$, for the field 
parallel to the [100] axis. 
Solid, dashed, and dash-dotted lines show the results for high carrier densities, 
$n=0.12$, $0.15$, and $0.2$, respectively. 
The dotted line is obtained in the crossover region, $n=0.1$, while dash-two-dotted line 
assumes a low carrier density, $n=0.06$. Fermi surfaces mainly consist of 
the d$_{\rm xy}$ orbital (d$_{\rm yz}$/d$_{\rm zx}$ orbitals) in the low (high) carrier density region. 
The other parameters are the same as those in Fig.~1. 
(b) Carrier density dependence of $\mu_{\rm B} H_{\rm c2}/T_{\rm c}$ at the lowest temperature 
$T/T_{\rm c}=0.05$ [circles]. 
We also show the results for $(t_{\rm odd}, \lambda) = (0.25,0)$ [squares], $(0,0.2)$ [diamonds], 
and $(0,0)$ [pluses] for comparison. 
}
\label{Fig.3}
\end{figure}

On the other hand, the Rashba spin-orbit coupling arising from the combination of 
$t_{\rm odd}$ and $\lambda$ leads to an intriguing superconducting phase in the magnetic field. 
Figure~2(a) shows the phase diagram against temperatures and magnetic fields 
for various carrier densities. 
We see an extraordinarily large normalized upper critical field, 
$\mu_{\rm B} H_{\rm c2}/T_{\rm c} > 9$, beyond the Pauli-Clogston-Chandrasekar limit, 
$\mu_{\rm B} H_{\rm c2}/T_{\rm c} = 1.25$,~\cite{Clogston} around $n = 0.12 - 0.15$. 
It has been shown that the upper critical field is enhanced by the Rashba spin-orbit 
coupling~\cite{Frigeriprl}, 
but that the enhancement is minor in the canonical Rashba-type non-centrosymmetric superconductors as 
$\mu_{\rm B} H_{\rm c2}/T_{\rm c} \approx 2$~\cite{YanaseCePt3Si-2}. 
We here find that the rather large enhancement of the upper critical field is caused by the synergy  
of the Rashba spin-orbit coupling and the orbital degree of freedom. 
Indeed, when we decrease the carrier density to $n <0.08$, the orbital degree of freedom is quenched 
and the upper critical field is suddenly decreased.

As shown in Fig.~2(b), the normalized upper critical field 
$\mu_{\rm B} H_{\rm c2}/T_{\rm c}$ shows a broad peak at approximately $n = 0.12$ and decreases 
with increasing carrier density for $n > 0.12$ except for a sharp enhancement at around $n=0.16$. 
The decrease in $\mu_{\rm B} H_{\rm c2}/T_{\rm c}$ is attributed to the decrease in Rashba 
spin-orbit coupling [see Fig.~1(b)]. 
A sharp peak at around $n=0.16$ is induced by the appearance of small Fermi surfaces 
around the $\Gamma$ point, that is, the Lifshitz transition. 
Because the g-factor of this band vanishes at ${\bm k}=(0,0)$ ($\Gamma$ point) 
in the presence of atomic LS coupling 
$\lambda$, Cooper pairing in the small Fermi surfaces is not disturbed by the magnetic field. 
Thus, a sharp enhancement of the normalized upper critical field, $\mu_{\rm B} H_{\rm c2}/T_{\rm c}$, is 
a signature of the Lifshitz transition. It will be interesting to look for this Lifshitz transition 
since the Class D topological superconducting phase is realized near the Lifshitz transition 
by applying a magnetic field~\cite{Sato-Fujimoto}. 
Since the renormalization of the g-factor is not due to the broken inversion symmetry, 
a sharp peak of $\mu_{\rm B} H_{\rm c2}/T_{\rm c}$ also appears for $(t_{\rm odd}, \lambda) = (0,0.2)$ 
[diamonds in Fig.~2(b)], for which the Lifshitz transition occurs at approximately $n=0.1$. 
Aside from this peak, a small upper critical field below the 
Pauli-Clogston-Chandrasekar limit is obtained when either $t_{\rm odd}$ or $\lambda$ is zero, because 
the Rashba spin-orbit coupling vanishes. 
As expected, the normalized upper critical field increases as we increase 
$t_{\rm odd}$ or $\lambda$. 
For instance, we obtain $\mu_{\rm B} H_{\rm c2}/T_{\rm c} \sim 4.9$ for $(t_{\rm odd}, \lambda) = (0.1,0.2)$ 
and  $n=0.12$, in agreement with the experimental result of STO/LAO interfaces~\cite{Shalom}.

\begin{figure}[ht]
\centering
\includegraphics[width=4.1cm,origin=c,keepaspectratio,clip]{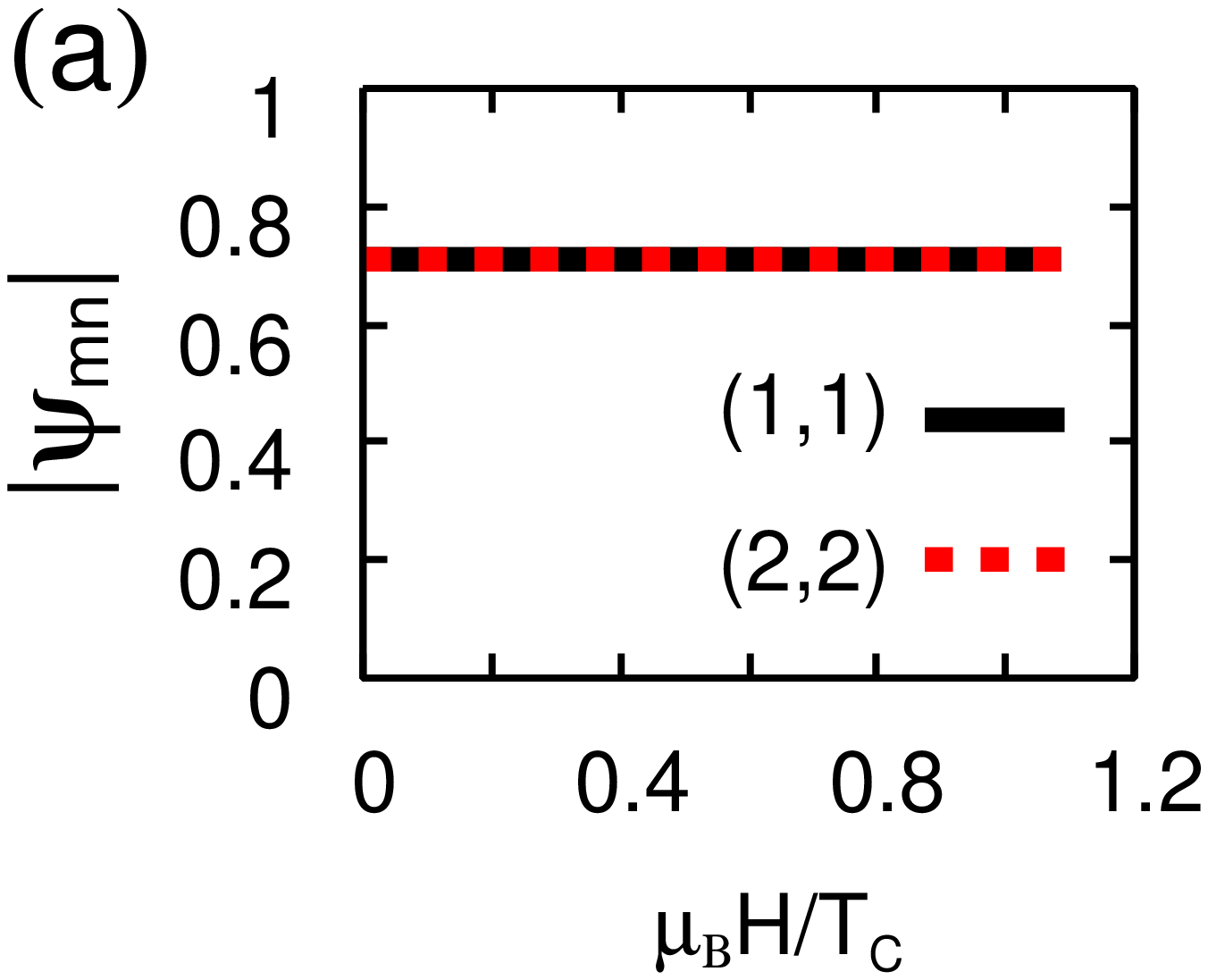}
\includegraphics[width=4.0cm,origin=c,keepaspectratio,clip]{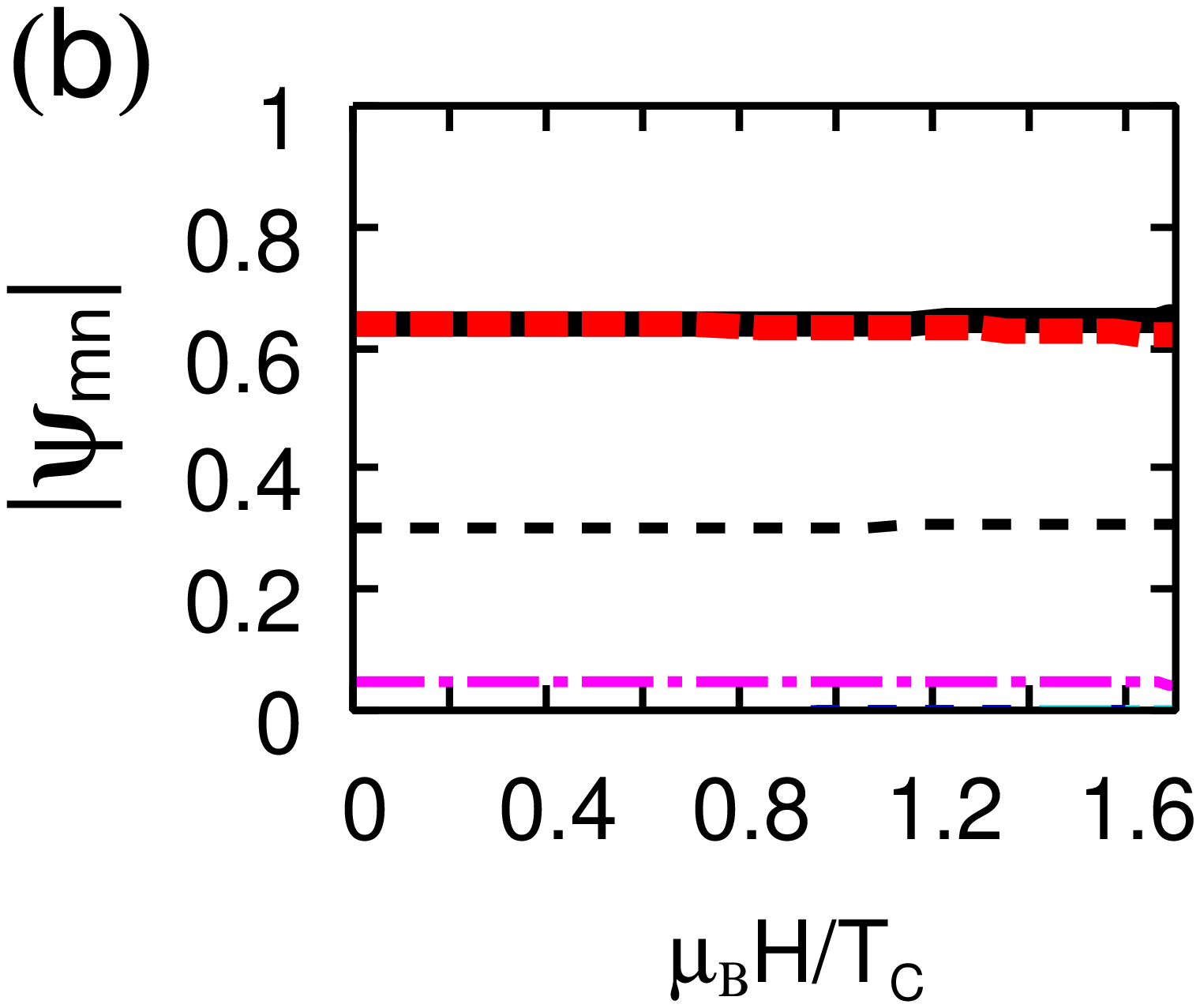}\\
\includegraphics[width=4.1cm,origin=c,keepaspectratio,clip]{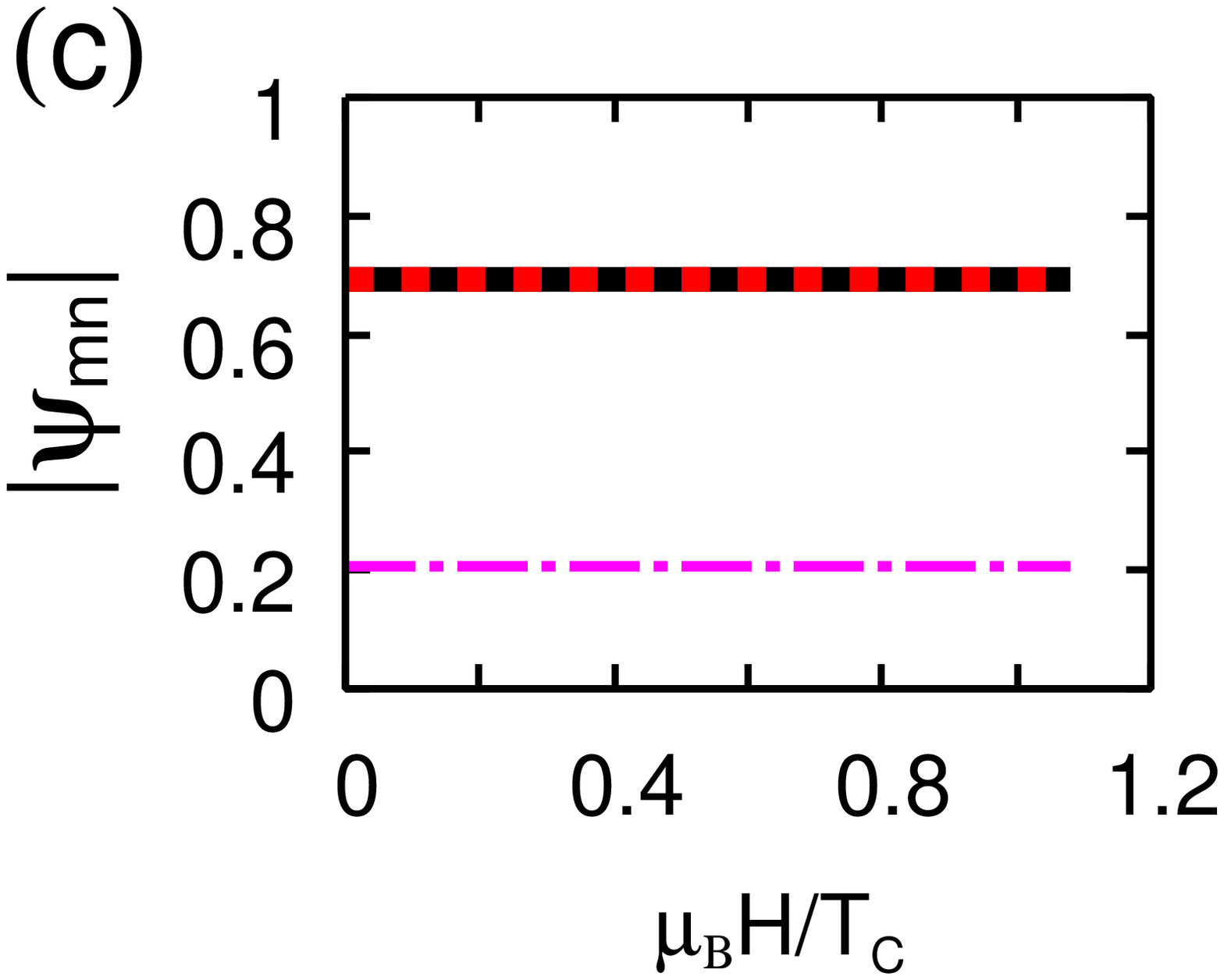}
\includegraphics[width=4.0cm,origin=c,keepaspectratio,clip]{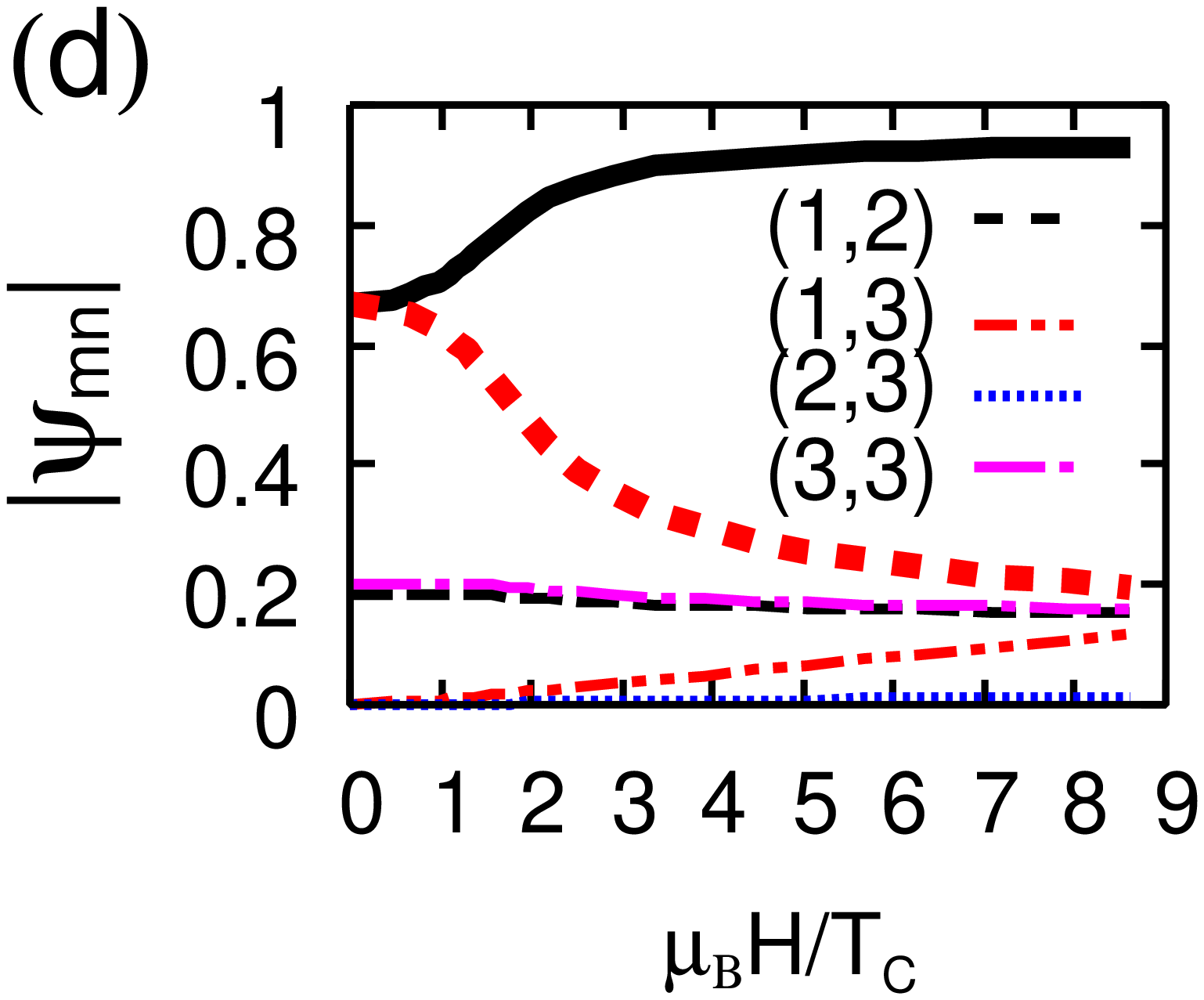}
\caption{(Color online) Magnetic field dependence of order parameters for $n=0.15$. 
We show the amplitude $|\psi_{mn}|$ at the transition temperature, 
which is proportional to the order parameters $|\Delta_{mn}|$ below $T_{\rm c}$. 
The main components are $|\psi_{11}|$ (thick solid line) and $|\psi_{22}|$ (thick dashed line). 
The other small components are shown by the thin lines, as described in Fig.~3(d). 
We assume $(t_{\rm odd}, \lambda) =$ (a) $(0,0)$, (b) $(0,0.2)$, (c) $(0.25,0)$, and (d) $(0.25,0.2)$. 
The other parameters are the same as those in Fig.~2. 
}
\label{Fig.4}
\end{figure}

In order to clarify the roles of the orbital degree of freedom, we show the magnetic 
field dependence of order parameters for a high carrier density, $n=0.15$. 
When both odd parity hybridization, $t_{\rm odd}$, and LS coupling, $\lambda$, are finite [Fig.~3(d)], 
the magnetic field along the {\it x}-axis substantially enhances the Cooper pairs of the d$_{\rm yz}$ orbital 
represented by $|\psi_{11}|$ while those of the d$_{\rm zx}$ orbital ($|\psi_{22}|$) are suppressed. 
This means that a quasi-one-dimensional superconducting state dominated by the d$_{\rm yz}$ orbital 
is stabilized in the magnetic field. Since this high-field superconducting phase is robust against 
the paramagnetic depairing effect, a large upper critical field is obtained, as shown in Fig.~2. 
It should be stressed that the Rashba spin-orbit coupling plays an essential role 
in stabilizing the quasi-one-dimensional superconducting phase. 
Indeed, we obtain a nearly isotropic two-dimensional superconducting phase 
with $|\psi_{11}| \sim |\psi_{22}|$ when either the odd parity hybridization $t_{\rm odd}$  
or the LS coupling $\lambda$ is zero [Figs.~3(a)-3(c)].

\begin{figure}[ht]
\centering
\includegraphics[width=5.5cm,origin=c,keepaspectratio,clip]{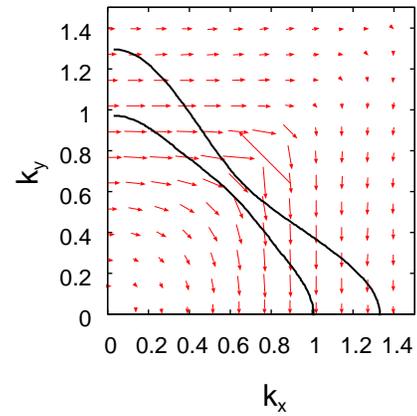}
\hspace{0.0cm}
\label{Fig.5}
\caption{(Color online) 
g-vector of the lowest band ($l=1$), $\g_1({\bm k})$, which is defined in eq.~(18). 
Arrows show the direction of the g-vector; the length of arrows is proportional to 
the amplitude of the g-vector. Solid lines show the Fermi surfaces for $n=0.15$. 
The other parameters are the same as those in Fig.~1. 
}
\end{figure}

We here illustrate why the quasi-one-dimensional superconducting phase is protected against the 
paramagnetic depairing effect. 
For this purpose, we derive the Rashba spin-orbit coupling in the band basis as 
we have performed in ref.~29.  
We reduce the single-particle Hamiltonian $H_{\rm 0}$ to the three-band model as 
$H_{\rm 0} = H_{\rm band} + H_{\rm ASOC}$, where 
\begin{eqnarray}
\label{band-basis}
&& \hspace*{-10mm}  
H_{\rm band} =\sum_{l=1}^3 \sum_{{\bm k} s} \xi_{l}({\bm k}) \ a_{{\bm k}, l s}^{\dag} \ a_{{\bm k}, l s}, 
\\ && \hspace*{-10mm}  
H_{\rm ASOC} =\sum_{l=1}^3 \sum_{{\bm k}}  \g_{l}({\bm k}) \cdot \SS_{l}({\bm k}),  
\end{eqnarray}
and $\xi_{l}({\bm k}) = \left(E_{2l}({\bm k}) + E_{2l-1}({\bm k})\right)/2$. 
The Rashba spin-orbit coupling of the $l$-th band is represented by the g-vector 
\begin{eqnarray}
\label{g-vector}
&& \hspace{-10mm}
\g_l({\bm k}) = \left(E_{2l}({\bm k})-E_{2l-1}({\bm k})\right) \SS_{2l}^{\rm \ av}({\bm k}) 
/|\SS_{2l}^{\rm \ av}({\bm k})|,  
\end{eqnarray} 
whose direction is obtained by calculating the average 
$\SS_{j}^{\rm \ av}({\bm k}) = \langle \sum_{m}\sum_{ss'} 
{\bm \sigma}_{ss'} c_{{\bm k}, \, m s}^{\dag}c_{{\bm k}, \, m s'} \rangle_j$ for the $j$-th eigenstate. 
Figure~4 shows the g-vector $\g_1({\bm k})$ and the Fermi surfaces for $n=0.15$.  
It is shown that the momentum dependence of the g-vector is quite different from an 
often-assumed form, $\g({\bm k}) = \alpha \left(\sin k_{\rm y}, - \sin k_{\rm x}, 0\right)$. 
This is the characteristic property of orbitally degenerate non-centrosymmetric 
systems~\cite{YanaseSr2RuO4_001}. 
In the case of STO heterostructures, 
the g-vector is nearly parallel to the {\it y}-axis for $k_{\rm x} > k_{\rm y}$, 
while it is almost along the {\it x}-axis for $k_{\rm x} < k_{\rm y}$. 
The quasiparticles mainly consist of the d$_{\rm yz}$ orbital (d$_{\rm zx}$ orbital) for the former (latter). 
Since the Cooper pairing is disturbed by the paramagnetic depairing effect when the g-vector is 
parallel to the magnetic field, the field along the $x$-axis suppresses the Cooper pairs of the  
d$_{\rm zx}$ orbital. 
On the other hand, the Cooper pairs formed by the d$_{\rm yz}$ orbital are protected 
by the g-vector nearly perpendicular to the magnetic field. 
In this way, the quasi-one-dimensional superconducting phase is stabilized by the orbital degree 
of freedom so as to avoid the paramagnetic depairing effect. This is an intuitive explanation for 
the large upper critical field shown in Fig.~2.

Finally, we discuss experimental results of the superconducting phase in STO/LAO interfaces. 
The superconducting transition temperature shows a non-monotonic carrier density 
dependence~\cite{Caviglia} and its peak at around $n = 2 \times 10^{13}$cm$^{-2}$ 
coincides with the crossover from d$_{\rm xy}$-orbital-dominated Fermi surfaces 
to d$_{\rm yz}$/d$_{\rm zx}$-orbital-dominated Fermi surfaces~\cite{Joshua}. 
The Rashba spin-orbit coupling seems to have the maximum amplitude  
in the crossover region~\cite{Shalom,Caviglia-2}, consistent with the 
three-orbital tight-binding model adopted in this study. 
Interestingly, a large upper critical field, $\mu_{\rm B} H_{\rm c2}/T_{\rm c} \approx 4.2$, beyond the 
Pauli-Clogston-Chandrasekar limit has been reported for a 
high carrier density $n = 3 \times 10^{13}$cm$^{-2}$ close to the crossover~\cite{Shalom}. 
The decrease in the normalized upper critical field $\mu_{\rm B} H_{\rm c2}/T_{\rm c}$ 
with increasing carrier density was also observed for  $n > 3 \times 10^{13}$cm$^{-2}$~\cite{Shalom}. 
These behaviors are consistent with our finding in Fig.~2, although the signature of Lifshitz transition 
has not been found. 
This agreement with experimental results indicates that the quasi-one-dimensional 
superconducting phase is realized in the STO/LAO interfaces with high carrier densities. 
In contrast to the theoretical proposal for a helical superconducting phase with a finite 
total momentum of Cooper pairs~\cite{Michaeli}, 
a large upper critical field is attributed to the entanglement of orbitals and spins 
in our three-orbital model. 
Indeed, we confirmed that the finite total momentum of Cooper pairs, namely, the finite ${\bm q}$ 
in the T-matrix, hardly changes our results. 
The coexistence of superconductivity and ferromagnetism~\cite{Dikin,Li,Bert,Ariando} 
may also be attributed to the quasi-one-dimensional superconducting phase protected against 
spin polarization. 
We would like to stress that such a spin-polarized superconducting state is hardly 
stabilized in the multiband models,~\cite{Fisher,Caprara} which phenomenologically assume 
the Rashba spin-orbit coupling and neglect the orbital degree of freedom. 
Our proposal for the quasi-one-dimensional superconducting phase can be verified by experiments using a 
tilted magnetic field. For instance, a vortex lattice structure elongated along the [010] axis will be 
observed in the field slightly tilted from the [100] axis to the [001] axis. 
As for a quantitative discussion, the crossover between low and high carrier density regions 
occurs in our model at around $n=0.08$, which corresponds to a carrier density of 
$n = 5 \times 10^{13}$ cm$^{-2}$. 
This is in reasonable agreement with experimental carrier density of $n = 2 \times 10^{13}$ cm$^{-2}$~\cite{Joshua}, 
and a discrepancy probably arises from our inexact choice of tight-binding parameters. 
Note that a large upper critical field has been observed in $\delta$-doped 
STO thin films~\cite{Kim-2}. Although the global inversion symmetry is not broken in this system, 
surface Rashba spin-orbit couplings play a similar role to the spin-orbit coupling in this study, 
as demonstrated for locally non-centrosymmetric superconductors~\cite{Maruyama}.

In summary, we studied the superconductivity in the two-dimensional electron systems formed at 
the STO/LAO interface and STO surface. We analyzed the three-orbital model taking into account 
$t_{\rm 2g}$ orbitals of Ti ions, and found that an unconventional structure of Rashba 
spin-orbit coupling arises from the orbital degeneracy and protects the quasi-one-dimensional 
superconducting phase against the paramagnetic depairing effect. 
The orbital degree of freedom plays an essential role in the response to the magnetic field and 
leads to a large upper critical field. 
The peak of the upper critical field as a function of carrier density coincides with the crossover 
from d$_{\rm xy}$-orbital-dominated Fermi surfaces to d$_{\rm yz}$/d$_{\rm zx}$-orbital-dominated Fermi surfaces. 
These observations provide a systematic understanding of superconducting properties 
at the STO/LAO interface.

The authors are grateful to S. Fujimoto and T. Shishido for fruitful discussions. 
This work was supported by KAKENHI (Grant Nos. 25103711, 24740230, and 23102709), 
and by a Grant for the Promotion of Niigata University Research Projects. 



\begin{thebibliography}{10}


\bibitem{Ohtomo}
A. Ohtomo and H. Y. Hwang: Nature {\bf 427} (2004) 423.

\bibitem{Biscaras}
J. Biscaras, N. Bergeal, A. Kushwaha, T. Wolf, A. Rastogi, R.C. Budhani, and J. Lesueur: 
Nat. Commun. {\bf 1} (2010) 89. 

\bibitem{Ueno}
K. Ueno, S. Nakamura, H. Shimotani, A. Ohtomo, N. Kimura, T. Nojima, 
H. Aoki, Y. Iwasa, and M. Kawasaki: 
Nat. Mater. {\bf 7} (2008) 855.

\bibitem{Kozuka}
Y. Kozuka, M. Kim, C. Bell, B. G. Kim, Y. Hikita, and H. Y. Hwang: 
Nature {\bf 462} (2009) 487. 


\bibitem{Reyen}
N. Reyren, S. Thiel, A. D. Caviglia, L. Fitting Kourkoutis, G. Hammerl, C. Richter, C. W. Schneider, T. Kopp, A.-S. R$\rm{\ddot{u}}$etschi, D. Jaccard, M. Gabay, D. A. Muller, J.-M. Triscone, and J. Mannhart: 
Science {\bf 317} (2007) 1196.


\bibitem{Brinkman}
A. Brinkman, M. Huijben, M. van Zalk, J. Huijben, U. Zeitler, J. C. Maan, W. G. van der Wiel, 
G. Rijnders, D. H. A. Blank, and H. Hilgenkamp: 
Nat. Mater. {\bf 6} (2007) 493.



\bibitem{Dikin}
D. A. Dikin, M. Mehta, C. W. Bark, C. M. Folkman, C. B. Eom, and V. Chandrasekhar: 
Phys. Rev. Lett. {\bf 107} (2011) 056802.

\bibitem{Li}
L. Li, C. Richter, J. Mannhart, and R. C. Ashoori: 
Nat. Phys. {\bf 7} (2011) 762.

\bibitem{Bert}
J. A. Bert, B. Kalisky, C. Bell, M. Kim, Y. Hikita, H. Y. Hwang, and K. A. Moler: 
Nat. Phys. {\bf 7} (2011) 767.

\bibitem{Ariando}
Ariando, X. Wang, G. Baskaran, Z. Q. Liu, J. Huijben, J. B. Yi, A. Annadi, A. Roy Barman, A. Rusydi, 
S. Dhar, Y. P. Feng, J. Ding, H. Hilgenkamp, and T. Venkatesan: 
Nat. Commun. {\bf 2} (2011) 188.


\bibitem{Caviglia}
A. D. Caviglia, S. Gariglio, N. Reyren, D. Jaccard, T. Schneider,
M. Gabay, S. Thiel, G. Hammerl, J. Mannhart, and J.-M. Triscone: 
Nature {\bf 456} (2008) 624.



\bibitem{Bell}
C. Bell, S. Harashima, Y. Kozuka, M. Kim, B. G. Kim, Y. Hikita, and H. Y. Hwang: 
Phys. Rev. Lett. {\bf 103} (2009) 226802.


\bibitem{Shalom}
M. Ben Shalom, M. Sachs, D. Rakhmilevitch, A. Palevski, and Y. Dagan: 
Phys. Rev. Lett. {\bf 104} (2010) 126802. 


\bibitem{Caviglia-2}
A. D. Caviglia, M. Gabay, S. Gariglio, N. Reyren, C. Cancellieri, and J.-M. Triscone: 
Phys. Rev. Lett. {\bf 104} (2010) 126803. 



\bibitem{Michaeli}
K. Michaeli, A. C. Potter, and P. A. Lee: 
Phys. Rev. Lett. {\bf 108} (2012) 117003. 


\bibitem{Springer}
{\it Non-Centrosymmetric Superconductors: Introduction and Overview},
ed. by E. Bauer and M. Sigrist (Springer-Verlag, Berlin, 2012). 


\bibitem{Banerjee}
S. Banerjee, O. Erten, and M. Randeria: 
arXiv:1303.3275. 


\bibitem{Santander-Sryo}
A. F. Santander-Syro, O. Copie, T. Kondo, F. Fortuna, S. Pailh${\rm \grave{e}}$s, R. Weht, X. G. Qiu, 
F. Bertran, A. Nicolaou, A. Taleb-Ibrahimi, P. Le F$\rm{\grave{e}}$vre, G. Herranz, M. Bibes, N. Reyren, 
Y. Apertet, P. Lecoeur, A. Barth$\rm{\acute{e}}$l$\rm{\acute{e}}$my, and M. J. Rozenberg
Nature {\bf 469} (2011) 189. 

\bibitem{Joshua}
A. Joshua, S. Pecker, J. Ruhman, E. Altman, and S. Ilani: 
Nat. Commun. {\bf 3} (2012) 1129. 


\bibitem{Popovic}
Z. S. Popovi${\rm \acute{c}}$, S. Satpathy, and R. M. Martin: 
Phys. Rev. Lett. {\bf 101} (2008) 256801. 


\bibitem{Pentcheva}
R. Pentcheva and W. Pickett: 
Phys. Rev. B {\bf 78} (2008) 205106. 


\bibitem{Delugas}
P. Delugas, A. Filippetti, V. Fiorentini, D. I. Bilc, D. Fontaine, and P. Ghosez: 
Phys. Rev. Lett. {\bf 106} (2011) 166807.


\bibitem{Khalsa}
G. Khalsa and A. H. MacDonald: Phys. Rev. B {\bf 86} (2012) 125121. 


\bibitem{Mizohata}
Y. Mizohata, M. Ichioka, and K. Machida: 
Phys. Rev. B {\bf 87} (2013) 014505. 

\bibitem{Fernandes}
R. M. Fernandes, J. T. Haraldsen, P. W${\rm\ddot{o}}$lfle, and A. V. Balatsky: 
Phys. Rev. B {\bf 87} (2013) 014510. 


\bibitem{Hirayama}
M. Hirayama, T. Miyake, and M. Imada: 
J. Phys. Soc. Jpn. {\bf 81} (2012) 084708. 


\bibitem{Zhong}
Z. Zhong, A. T${\rm \acute{o}}$th, and K. Held: Phys. Rev. B {\bf 87} (2013) 161102.  

\bibitem{Khalsa2}
G. Khalsa, B. Lee, and A. H. MacDonald: arXiv:1301.2784.




\bibitem{YanaseSr2RuO4_001}
Y.~Yanase: J. Phys. Soc. Jpn. {\bfseries 82} (2013) 044711; 
Y. Yanase and H. Harima: Kotai-Butsuri {\bf 47} (2012) No.~3, 1 [in Japanese]. 


\bibitem{Nayak}
Y. Kim, R. M. Lutchyn, and C. Nayak: arXiv:1304.0464. 



%
%
%

\bibitem{YanaseCePt3Si} 
Y. Yanase and M. Sigrist: 
J. Phys. Soc. Jpn. {\bf 77} (2008) 124711. 

\bibitem{Nagano-Shishidou-Oguchi}
M. Nagano, A. Kodama, T. Shishidou, and T. Oguchi: 
J. Phys: Condens. Matter {\bf 21} (2009) 064239. 



\bibitem{Binnig}
G. Binnig, A. Baratoff, H. E. Hoenig, J. G. Bednorz: 
Phys. Rev. Lett. {\bf 45} (1980) 1352. 




\bibitem{Yada}
K. Yada, S. Onari, Y. Tanaka, and J. Inoue: 
Phys. Rev. B {\bf 80} (2009) 140509. 

\bibitem{Bert2}
J. A. Bert, K. C. Nowack, B. Kalisky, H. Noad, J. R. Kirtley, C. Bell, H. K. Sato, M. Hosoda, 
Y. Hikita1, H. Y. Hwang, and K. A. Moler: 
Phys. Rev. B {\bf 86} (2012) 060503(R). 



\bibitem{Aoyama}
K. Aoyama and M. Sigrist: Phys. Rev. Lett. {\bf 109} (2012) 237007.


\bibitem{Clogston}
B. S. Chandrasekhar: Appl. Phys. Lett. {\bf 1} (1962) 7; 
A. M. Clogston: Phys. Rev. Lett. {\bf 9} (1962) 266. 


\bibitem{Frigeriprl}
P.~A. Frigeri, D.~F. Agterberg, A. Koga, and M. Sigrist:  
Phys. Rev. Lett. {\bf 92} (2004) 097001. 


\bibitem{YanaseCePt3Si-2}
Y.~Yanase and M.~Sigrist: J. Phys. Soc. Jpn. {\bfseries 76} (2007) 124709. 


\bibitem{Sato-Fujimoto}
M. Sato, Y. Takahashi, and S. Fujimoto: 
Phys. Rev. Lett. {\bf 103} (2009) 020401.  



\bibitem{Fisher}
M. H. Fischer, S. Raghu, and E.-A. Kim: 
New J. Phys. {\bf 15} (2013) 023022. 

\bibitem{Caprara}
S. Caprara, F. Peronaci, and M. Grilli: 
Phys. Rev. Lett. {\bf 109} (2012) 196401. 


\bibitem{Kim-2}
M. Kim, Y. Kozuka, C. Bell, Y. Hikita, and H. Y. Hwang: 
Phys. Rev. B {\bf 86} (2012) 085121. 


\bibitem{Maruyama}
D. Maruyama, M. Sigrist, and Y. Yanase: 
J. Phys. Soc. Jpn. {\bf 81} (2012) 034702. 


\end{thebibliography}
\end{document}